\documentclass[reprint,
superscriptaddress,
 amsmath,amssymb,
 aps,
 prl
]{revtex4-1}
\newcommand{\gdd}[0]{g_\mathrm{dd}}
\newcommand{\sro}[0]{{\sqrt{\rho}}}

\usepackage{graphicx}
\usepackage{xcolor}
\usepackage{hyperref}
\usepackage{dcolumn}
\usepackage{bm}

\begin{document}

\preprint{APS/123-QED}

\title{
Ultrawide dark solitons and droplet-soliton coexistence 
in a dipolar Bose gas\\with strong contact interactions
}

\author{Jakub Kopyci\'{n}ski}
\email{jkopycinski@cft.edu.pl}
\affiliation{
Center for Theoretical Physics, Polish Academy of Sciences, Al. Lotnik\'{o}w 32/46, 02-668 Warsaw, Poland
}
\author{Maciej \L ebek}
\affiliation{
Center for Theoretical Physics, Polish Academy of Sciences, Al. Lotnik\'{o}w 32/46, 02-668 Warsaw, Poland
}
\author{Wojciech G\'{o}recki}
\affiliation{Faculty of Physics, University of Warsaw, Pasteura 5, 02-093 Warsaw, Poland}
\author{Krzysztof Paw\l owski}
\affiliation{
Center for Theoretical Physics, Polish Academy of Sciences, Al. Lotnik\'{o}w 32/46, 02-668 Warsaw, Poland
}

\date{\today}

\begin{abstract}
We look into dark solitons in a quasi-1D dipolar Bose gas and in a quantum droplet. We derive the analytical solitonic solution of a Gross-Pitaevskii-like equation accounting for beyond mean-field effects.
The results show there is a certain critical value of the dipolar interactions, for which the width of a motionless soliton diverges. Moreover, there is a peculiar solution of the motionless soliton with a non-zero density minimum. We also present the energy spectrum of these solitons with an additional excitation subbranch appearing. Finally, we perform a series of numerical experiments revealing the coexistence of a dark soliton inside a quantum droplet.
\end{abstract}

\maketitle

\begin{figure}[t!]
\includegraphics[width=\linewidth]{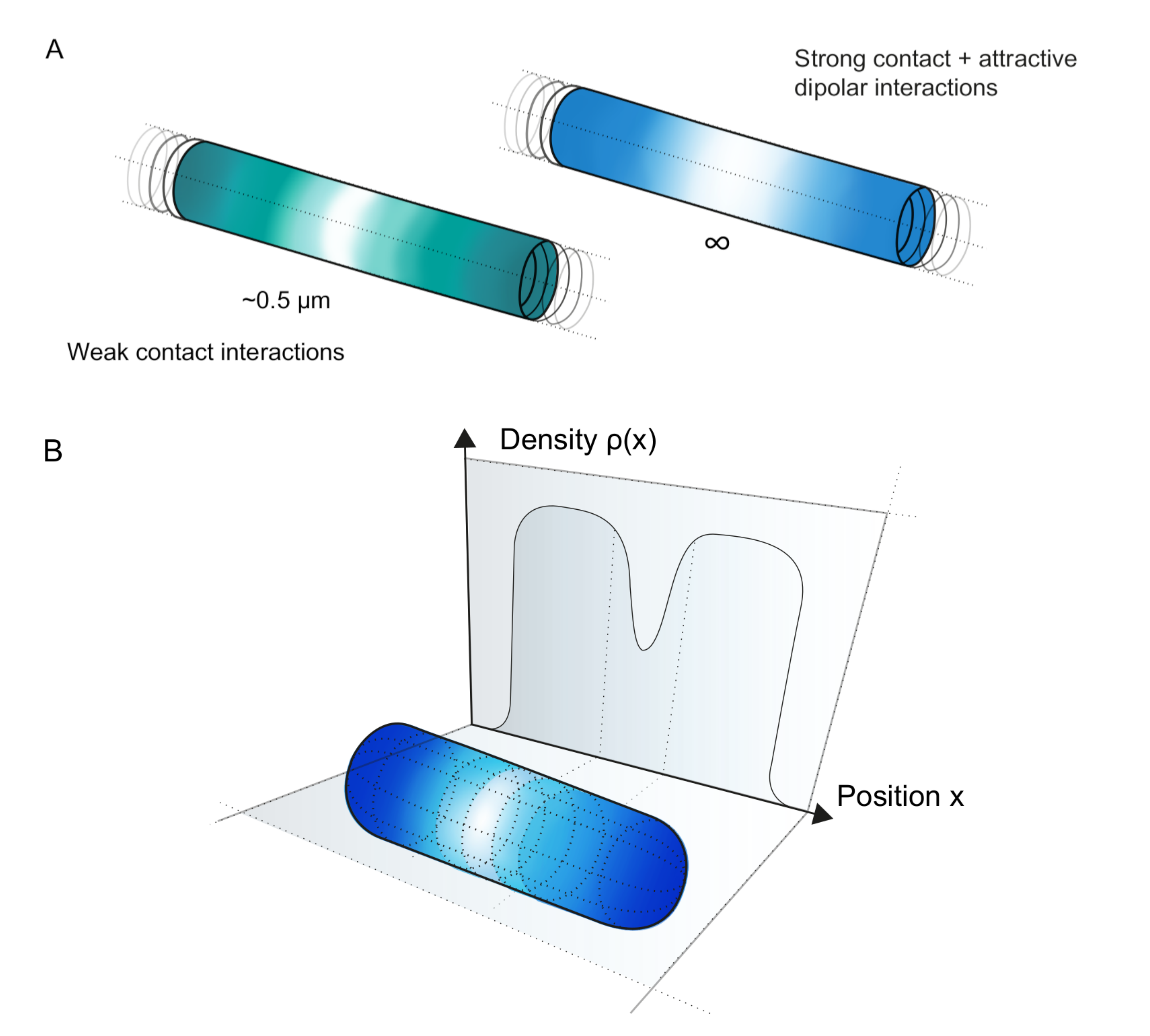}
\caption{\label{fig:graph_abs} Graphical abstract. A. We demonstrate that in a dipolar gas, there are solutions of infinitely wide dark solitons due to an interplay between short- and long-range interactions. B. Artistic vision of a dark soliton existing inside a quantum droplet.
}
\end{figure}
\section{Introduction}

In this Letter, we wish to address two vital topics  in the field of ultracold atoms: the investigation of quantum droplets, the subject of dipolar dark solitons.

The early experiments with attractive dipolar Bose-Einstein condensates (BEC) reported the gas collapse dubbed Bose-nova~\cite{Sackett_1999, Gerton_2000, Donley_2001, Roberts_2001}. Later on, opening doors to some new species with a high magnetic dipole moment~\cite{Griesmaier_2005, Aikawa_2012, Lu_2012} offered more possibilities. Apart from 
the next evidence of
the collapse~\cite{Lahaye_2008, Metz_2009, Lahaye_2009}, a state with a broken symmetry was unexpectedly brought to light~\cite{Kadau_2016}. Subsequently, the novel states of matter -- quantum droplets~\cite{Schmitt_2016, Cabrera_2018, Semeghini_2018} and supersolids~\cite{Wenzel_2017, Sohmen_2021}, were observed in  dipolar systems and Bose-Bose mixtures.

The collapse-preventing mechanism in mixtures~\cite{Petrov_2015} and dipolar BECs~\cite{Ferrier-Barbut_2016} is due to the quantum fluctuations, not accounted for in the seminal Gross-Pitaevskii equation (GPE). 
Therefore, one may look for an extended Gross-Pitaevskii equation (eGPE). For instance, the Kolomeisky equation~\cite{Kolomeisky_2000} was proposed to describe the Tonks-Girardeau gas~\cite{Girardeau_1960}, but suffered an immediate criticism~\cite{Girardeau_2000}. Another example of an eGPE is the generalized nonlocal nonlinear Schr\"odinger equation~\cite{Wachtler_2016a, Wachtler_2016b, Bisset_2016} employing the Lee-Huang-Yang (LHY) correction~\cite{Lee_1957a, Lee_1957b}. 
%Such an approach, 
GPE extended by the LHY correction only, does not provide us with a quantitative agreement with quantum Monte Carlo predictions for strong interactions~\cite{Bottcher_2019, Parisi_2019}. 

The gap of strong interactions seems to be filled with 
%yet another GP-like equation, called
a hydrodynamic-based approach resulting in the Lieb-Liniger GPE (LLGPE), which we use in this Letter.
The equation in question was used to investigate BECs in Refs.~\cite{Dunjko_2001, Kim_Zubarev_2003,  Damski_2004, Damski_2006, Oehberg_2002,  Kopycinski_2022, Peotta_2014, Choi_2015, DePalo_2021}, including the prediction of dipolar quantum droplets in a quasi-1D configuration~\cite{Oldziejewski_2020}. 

Typically, these waves due their existence to nonlinear effects and have been already studied in various physical systems~\cite{Poy_2022, Chai_2020, Bersano_2018, Hoefer_2011, Theocharis_2010, Becker_2008,  Weller_2008, Rotschild_2006,  Kivshar_1998}, including BECs with contact interactions~\cite{Zakharov_1973, Jackson_1998, Sato_2016}, dipolar interactions~\cite{Pawlowski_2015, Beau_2022}, and also beyond the mean-field description~\cite{Kolomeisky_2000, Kopycinski_2022}. It was shown that in the TG gas, thanks to the Bose-Fermi mapping, one can find stable soliton-like solutions in many-body calculations~\cite{Tettamanti_2021}.

Dark solitons are routinely produced
via phase imprinting~\cite{Burger_1999, Denschlag_2000}, but they are predicted to appear spontaneously during heating as well~\cite{Karpiuk_2012}.
Unfortunately, the solitons are too narrow to be observed {\it in situ}, which makes an obstacle to investigate them properly.

The aim of this Letter is to find out whether or not dark solitons exist in a dipolar Bose gas with strong contact interactions and if such solitons can coexist with quantum droplets. As far as we are concerned, the necessary ingredients, namely strong contact interactions, quasi-1D geometry, and dipolar interactions, have been present in the experiment~\cite{Kao_2020}.

\section{Framework}

\subsection{\label{sec:NLLGPE}Nonlocal Lieb-Liniger Gross-Pitaevskii equation}
We study a dipolar Bose gas in a quasi-1D configuration, i.e.
with unconfined atoms of mass $m$
in the $x$ direction (box of a size $L$ with periodic boundary conditions or an infinite 
system), but tightly trapped 
in the transverse $y$ and $z$ directions. 
We assume a harmonic trap with frequency $\omega_\perp$ and introduce the aspect ratio $\sigma=l_\perp/L$ with $l_\perp=\sqrt{\hbar/m\omega_\perp}$. 
%Now, one can use the following form of the pseudo-wavefunction~\cite{Kolomeisky_2000, Kim_Zubarev_2003,  Oldziejewski_2020, Oehberg_2002}: $\Psi(\bm{r},  t)=\Phi(x,t)\psi_G(y)\psi_G(z)$, where $\psi_G(y)=(\pi l_\perp^2)^{-1/4}\exp(-y^2/(2l_\perp^2))$. 
The excitation energies in the perpendicular directions are assumed to be relatively large in comparison to those in the longitudinal one, therefore unoccupied and neglected here.

We consider a head-to-tail configuration of the dipoles. This implies the attractive character of the dipole-dipole interaction (DDI). The dipolar interaction coefficient $\gdd\equiv\frac{\mu_0 \mu_D^2}{2l_\perp^2}$ depends on the atom magnetic moment~$\mu_D$. 

To understand the dynamics of the system, we start with classical hydrodynamical Euler conservation equations 
\begin{subequations}
\begin{equation}
    \frac{\partial\rho}{\partial t}+\frac{\partial(\rho v)}{\partial x}=0,
\end{equation}
\begin{equation}
    \frac{\partial v}{\partial t}+v \frac{\partial v}{\partial x}=\frac{1}{m\rho}\frac{\partial P_{\rm LL}}{\partial x}+\mathrm{f},\label{eq:hydroB}
\end{equation}
\label{eq:hydro}
\end{subequations}
where $P_{\rm LL}$ is the pressure and $\mathrm{f}$ contains body accelerations. We assume further that all fields change sufficiently slow such that locally the gas remains in the ground state of the Lieb-Liniger model with energy $E_{\rm LL}$, and therefore $P_{\rm LL}=-\frac{\partial E_{\rm LL}}{\partial L}$ (which we calculate only in the thermodynamic limit~\cite{Mancarella_2014}), while dipolar magnetic forces are included in $\mathrm{f}$.

We introduce a pseudo-wavefunction  $\Phi(x,t)$, which has to obey the above set of hydrodynamic equations via the density field $\rho(x,t)=|\Phi(x,t)|^2$ and velocity field $v(x,t)=\frac{\hbar}{m} \partial_x [\arg\Phi(x,t)]$. This, up to the quantum pressure term, leads to the nonlocal LLGPE~\footnote{The explicit form of body acceleration in Eq.~\eqref{eq:hydroB} is given by  $\mathrm{f}(x,t)=-\frac{1}{m\rho(x,t)}\frac{\partial}{\partial x}\gdd\int\rho(x',t)v_\mathrm{dd}^\sigma(|x-x'|)\rho(x,t)dx'$.}:
\begin{eqnarray}
     i\hbar\partial_t \Phi(x,t)=\left(-\frac{\hbar^2}{2m}\partial_{xx} +\mu_\mathrm{LL}[g, |\Phi(x,t)|^2]\right)\Phi(x,t)\nonumber\\
     -\gdd\int|\Phi(x',t)|^2v_\mathrm{dd}^\sigma(|x-x'|)\Phi(x,t)dx',\quad
     \label{eq:full_NLLGPE}
\end{eqnarray}
where $g$ is the contact interaction coefficient, $v_\mathrm{dd}^\sigma(u)=\frac{1}{\sigma}v_\mathrm{dd}(u/\sigma)$ is the effective quasi-1D dipolar potential~\cite{Deuretzbacher_2010} with $v_\mathrm{dd}(u)=\frac{1}{4}\left[-2|u|+\sqrt{2\pi}(1+u^2)e^{u^2/2}\mathrm{Erfc}(|u|/\sqrt{2})\right]$ and $\mu_\mathrm{LL}$ is the  Lieb-Liniger chemical potential~\cite{Lieb_1963a}. The latter can be easily evaluated numerically basing on Refs.~\cite{Ristivojevic_2019, Lang_2017}. The parameter $\sigma$ acquires another meaning of the effective dipolar interaction range here.

For the weak contact interactions the subsequent terms of $\mu_\mathrm{LL}$ expanded in the Taylor series give rise to the GPE and eGPE (that is GPE with LHY term). Keeping in \eqref{eq:full_NLLGPE} the full $\mu_\mathrm{LL}$, without cutting the Taylor series, makes the equation useful for strong short-range interactions \cite{Kopycinski_2022} and a tool to study the 1D quantum droplets \cite{Oldziejewski_2020}.

\subsection{Approximation of infinitely strong contact interactions with zero-range dipolar ones}
Following the notation from the Lieb-Liniger model, we will use a dimensionless parameter $\gamma\equiv\frac{m}{\hbar^2}\frac{g}{\rho_0}$ to describe the contact interaction strength, where $\rho_0=N/L$ is the average gas density. Let us now consider Eq.~\eqref{eq:full_NLLGPE} in the limit of the infinite contact and zero-range dipolar interactions (i.e. $\gamma\to\infty$ and $\sigma\to0$):
\begin{eqnarray}
     i\hbar\partial_t \Phi(x,t)=\frac{\hbar^2}{2m}\left(-\partial_{xx} +\pi^2|\Phi(x,t)|^4\right)\Phi(x,t)\nonumber\\
     -\gdd|\Phi(x,t)|^2\Phi(x,t),\quad
     \label{eq:Malomed_Salerno_eqn}
\end{eqnarray}
which is the very equation from Ref.~\cite{Baizakov_2009}. As it can be immediately seen, the part of this equation responsible for the short-range interactions  becomes the same as in the Kolomeisky equation~\cite{Kolomeisky_2000}. On the other hand, the dipolar part acquires the nonlinear form known from the GPE. This approximation works well when the interaction range $\sigma$
is smaller than the typical length scale over which density can change, but still much larger than the average interparticle distance. According to Ref.~\cite{Pawlowski_2015}, the solitonic solution in the repulsive dipolar gas in this limit is convergent to the one in the gas with contact interactions only.

One of the
solutions of Eq.~\eqref{eq:Malomed_Salerno_eqn}, $\Phi_\mathrm{MS}(x)$ was found in Ref.~\cite{Baizakov_2009}. It was initially thought to be a family of bright solitons. However, as some solutions have a flat-top density profile and the energy linear in $N$, we refer to these objects as quantum droplets. Such crossovers, from a bright soliton to a quantum droplet, were seen in dipolar systems~\cite{Oldziejewski_2020} and mixtures~\cite{Cheiney_2018, Cui_2021}.

\section{Results}
\subsection{Speed of sound and stability of the constant density profile}
We introduce a dimensionless parameter $\gamma_\mathrm{dd}\equiv\frac{m}{\hbar^2}\frac{\gdd} {\rho_0}$ describing the dipolar interaction strength, mimicking the Lieb parameter $\gamma$.

First, we linearize Eq.~\eqref{eq:Malomed_Salerno_eqn}
for a homogenous system 
and solve Bogoliubov-de Gennes equations~\cite{Bogoliubov_1947} to get the excitation energy
$\epsilon(k)=\sqrt{\frac{\hbar^4k^4}{4m^2}+\hbar^2k^2\left(\frac{\hbar^2\pi^2\rho_0^2}{m^2}-\frac{\gdd\rho_0}{m}\right)}$ as a function of the wave 
vector $k$. For low momenta the spectrum is linear, i.e. it contains phonons with the speed of sound
$c=\sqrt{\frac{\hbar^2\rho_0^2}{m^2}\left(\pi^2-\gamma_\mathrm{dd}\right)}$.
When $\gamma_\mathrm{dd}>\pi^2$, the Bogoliubov excitation energy becomes complex for low momenta, manifesting the phonon instability in this region. Such an instability is present in a Bose gas with attractive interactions, where  the ground state breaks the translational symmetry to form a bright soliton~\cite{Kanamoto_2003}. In our case, however, the symmetry-broken ground state appears as soon as $\gamma_\mathrm{dd}>\frac{2}{3}\pi^2$ where the pressure $P=-\frac{\partial E_0}{\partial L}=\frac{\hbar^2\rho_0^3}{m}\left(\frac{\pi^2}{3}-\frac{\gamma_\mathrm{dd}}{2}\right)$~\cite{Kopycinski_2022}, with $E_0$ being the homogeneous state energy, becomes negative. Thus, pressure is the very parameter which determines the emergence of quantum droplets. 
\subsection{Dark soliton solution}

We look for a family of Eq.~\eqref{eq:Malomed_Salerno_eqn} solutions such that $\Phi_s(x,t)\equiv\psi(x-vt)\exp(-i\mu t/\hbar)$ and $\psi(\zeta)=\sqrt{\rho(\zeta)}e^{i\phi(\zeta)}$. Real-valued functions $\rho$ and $\phi$ are interpreted as the density and phase respectively. Parameter $v$ is the soliton velocity, $\zeta=x-vt$ is the comoving coordinate, and $\mu$ is the chemical potential.
When we apply $\Phi_s(x,t)$ to Eq.~\eqref{eq:Malomed_Salerno_eqn},

we obtain the soliton density and phase profiles
\begin{subequations}
\label{eq:solutions}
	\begin{equation}
  		 \rho(\zeta)=\rho_\infty-\frac{(\rho_\infty-\rho_\mathrm{min})(1+D)}{1+D\cosh(W\zeta)}, 
    	\label{eq:density_sol}
	\end{equation}
    \begin{equation}
	    \begin{aligned}
  		\phi(\zeta)=\frac{2mv(D+1)(\frac{\rho_\mathrm{min}}{\rho_\infty}-1)}{\hbar DW\sqrt{1-a^2}}\\
  		\times\arctan\left(\frac{(a-1)\tanh\left(\frac{W\zeta}{2}\right)}{\sqrt{1-a^2}}\right),
    	\label{eq:phase_sol}
        \end{aligned}	
    \end{equation}
\end{subequations}
where $a\equiv \frac{\rho_\mathrm{min}}{\rho_\infty}+\frac{\rho_\mathrm{min}}{D\rho_\infty}-1$, $D\equiv\frac{\rho_\mathrm{min}-\rho_1}{2\rho_\infty-\rho_1-\rho_\mathrm{min}}$, and $W\equiv 2\sqrt{\frac{\pi^2}{3}(\rho_\infty-\rho_\mathrm{min})(\rho_\infty-\rho_1)}$. Constants  $\rho_\mathrm{min}=\frac{3m\gdd}{2\hbar^2\pi^2}-\rho_\infty+\frac{\sqrt{\Delta}}{2}$ with $\Delta=\left(2\rho_\infty-\frac{3m\gdd}{\hbar^2\pi^2}\right)^2+\frac{12m^2v^2}{\hbar^2\pi^2}$ is interpreted as the soliton density minimum and $\rho_\infty$ is the background density ($\lim_{\zeta\to\pm\infty}\rho(\zeta)=\rho_\infty$). There is no clear interpretation for $\rho_1=\frac{3m\gdd}{2\hbar^2\pi^2}-\rho_\infty-\frac{\sqrt{\Delta}}{2}$, though. In the thermodynamic limit $\rho_\infty=\rho_{0}$.

\begin{figure}[h!]
\includegraphics[width=\linewidth]{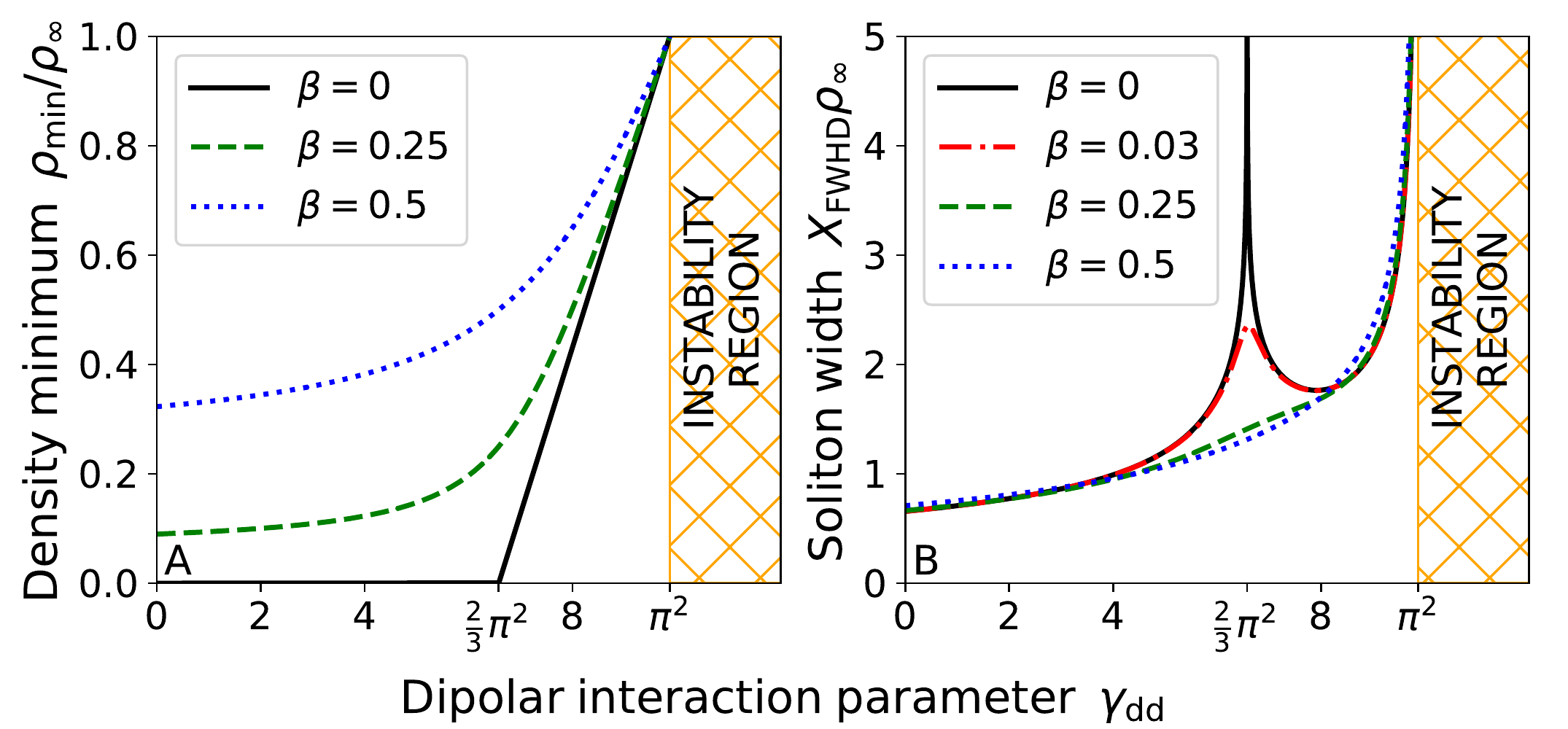}
\caption{\label{fig:densmin+fwhd}A. Soliton density minima $\rho_\mathrm{min}$ as functions of the dipolar interaction strength $\gamma_\mathrm{dd}$ for different relative velocities~$\beta$.\\
B. Soliton full width at half depth $X_\mathrm{FWHD}$ as functions of the dipolar interaction strength $\gamma_\mathrm{dd}$ for different relative velocities~$\beta$. }
\end{figure}

One can notice an interesting feature analysing solely the soliton density minimum as a function of the dipolar interaction strength $\gamma_\mathrm{dd}$. As we can see it in Fig.~\ref{fig:densmin+fwhd}A, the density of the motionless soliton ($\beta\equiv v/c=0$) above $\gamma_\mathrm{dd}>\frac{2}{3}\pi^2$ is not vanishing. Moreover, the phase of such a solution is constant.

Another vital property of the soliton in question is its full width at half depth $X_\mathrm{FWHD}=\frac{2}{W}\mathrm{arccosh}\left(\frac{1+2D}{D}\right)$ such that $\rho\left(\frac{X_\mathrm{FWHD}}{2}\right)=\frac{\rho_\infty+\rho_\mathrm{min}}{2}$. Fig.~\ref{fig:densmin+fwhd}B shows the motionless soliton width diverges logarithmically when $\gamma_\mathrm{dd}\to \frac{2}{3}\pi^2$. Even if $0 < \beta \ll 1$, we can see the soliton size at $\gamma_\mathrm{dd}=\frac{2}{3}\pi^2$ is larger than the interparticle distance $1/\rho_\infty$ and might be easier to detect in the experiment.

The soliton width also diverges when $\gamma_\mathrm{dd}\to\pi^2$ but in that case $X_\mathrm{FWHD}\propto |\gamma_\mathrm{dd}-\pi^2|^{-\nu}$ with the critical exponent $\nu=1/2$.

One can also notice the phase difference $\Delta\phi_\mathrm{max}=\max \left[\lim_{\zeta\to\infty}\phi(\zeta)-\lim_{\zeta\to-\infty}\phi(\zeta)\right]$ is equal to $\pi$ when $\gamma_\mathrm{dd} < \frac{2}{3}\pi^2$, but it is smaller than $\pi$ otherwise~\footnote{In addition to this, when $\gamma_\mathrm{dd}>\frac{2}{3}\pi^2$, there is more than one state corresponding to the same phase imprint $\Delta\phi$. For instance, when $\Delta\phi=0$ we can have either a motionless soliton or a flat density profile.}. We hypothesize a phase imprint of $\Delta \phi>\Delta\phi_\mathrm{max}$ on a droplet may lead to its splitting rather than a soliton formation. We investigate this case later in this Letter.

\subsection{Dispersion relation}

\begin{figure}[h!]
\includegraphics[width=\linewidth]{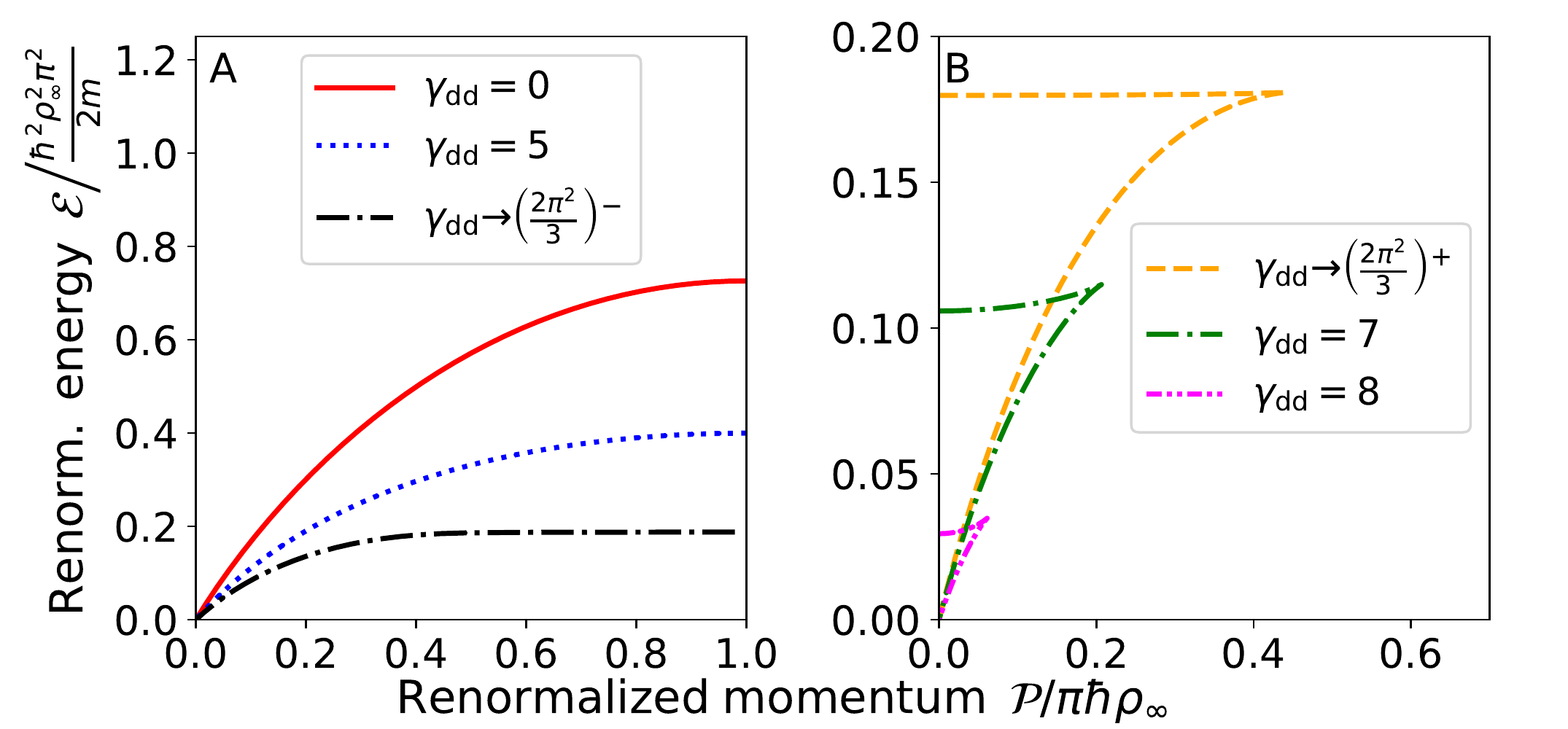}
\caption{\label{fig:disp} A. Soliton dispersion relation $\mathcal{E}(\mathcal{P})$ for $\gamma_\mathrm{dd}<\frac{2}{3}\pi^2$. \\
B. Soliton dispersion relation $\mathcal{E}(\mathcal{P})$ for $\gamma_\mathrm{dd}>\frac{2}{3}\pi^2$. 
}
\end{figure}

We calculate and show in Fig.~\ref{fig:disp} the dispersion relation $\mathcal{E}(\mathcal{P})$~\footnote{Both energy and momentum demand proper renormalizations according to Refs.~\cite{Kivshar_1998, Kolomeisky_2000}. See Supplemental Material at [URL will be inserted by publisher] for the details.} of the soliton renormalized energy $\mathcal{E}$ and momentum $\mathcal{P}$.

In the range $0 \leqslant \gamma_\mathrm{dd} < \frac{2}{3}\pi^2$, the dispersion relation behaves qualitatively the same as the one coming from the Kolomeisky solution ($\gamma_\mathrm{dd}=0$)~\cite{Kolomeisky_2000}, whereas, when $\gamma_\mathrm{dd}>\frac{2}{3}\pi^2$, we observe a new subbranch formed. From now on, we refer to these solutions as anomalous solitons.

Obviously, it may be more difficult to phase imprint the anomalous solitons just beacause their excitation energy is higher than the one corresponding to the solution with the same phase difference $\Delta\phi$, but situated on the lower subbranch. We cannot also perform a phase imprint of a motionless anomalous soliton.

Another problem may be encountered when trying to calculate the effective soliton mass $m^*=\hbar^2\left(\frac{d^2\mathcal{E}}{dk^2}\right)^{-1}$. It is not well defined due to the presence of a cusp in the spectrum. 

\subsection{\label{sec:ph_impr}Dark soliton generation inside a quantum droplet}

Last but not least, we want to find out whether or not dark solitons may coexist with quantum droplets. We prepared two series of numerical experiments using the MUDGE toolkit~\footnote{MUDGE is a toolkit created to simulate 1D Bose systems and uses the W-DATA format~\cite{WSLDAToolkit, Bulgac_2014, Wlazlowski_2018}. See Supplemental Material at [URL will be inserted by publisher] for code details.}. Assuming the soliton width is much smaller than the droplet size, one can treat the droplet bulk density $\rho_\mathrm{eq}$ as $\rho_\infty$ and expect that after a phase imprint of $\Delta\phi$ phase difference, the imprinted state is similar to the analytical solitonic solution.

Note that then, due to the interplay between the short-range and dipole interaction, consequently rescaled interaction parameter $\frac{\rho_0}{\rho_\mathrm{eq}}\gamma_\mathrm{dd}$ does not depend on the actual value of $\gamma_\mathrm{dd}$ and is equal $\frac{2}{3} \pi^2\left[1-\text{sech}\left(\pi\sqrt{N/3}\right)\right]^{-1}$~\footnote{See Supplemental Material at [URL will be inserted by publisher] for the details}, which corresponds to the left edge of the anomalous region in Fig.~\ref{fig:densmin+fwhd}.

\begin{figure}[h!]
\includegraphics[width=\linewidth]{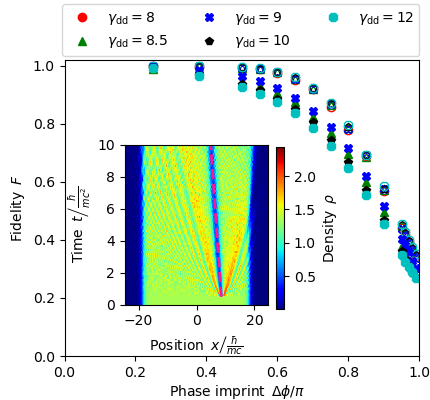}
\caption{\label{fig:fidelities} Fidelity $F$ between a state that appears dynamically after  phase-imprinting and the analytical solitonic solution. Empty markers: $\gamma\to\infty$ and $\sigma=0$, filled markers: $\gamma=50$ and $\sigma=0.05$\\
Inset: Evolution of a quantum droplet given by Eq.~\eqref{eq:Malomed_Salerno_eqn} with dipolar interaction strength $\gamma_\mathrm{dd}=9$ after a phase imprint $\Delta\phi=\pi/2$ at $t\approx \frac{mc^2}{\hbar}$. The dashed line  marks a trajectory  of an object moving with velocity $\beta=0.3684(4)$, which was predicted from the shape of the soliton 
in accordance with Fig.~\ref{fig:densmin+fwhd}. The visible emerging high-density structures are the shock waves induced in the phase-imprinting process.
}
\end{figure}

We use the fidelity $F=|\langle \psi^\sigma_\mathrm{num}|\psi\rangle|^2$ between the numerically evaluated phase-imprinted state $\psi^\sigma_\mathrm{num}$ and the solitonic solution $\psi=\sqrt{\rho}e^{i\phi}$ with density and phase given by Eqs.~\eqref{eq:solutions}. We have $\sigma=0$ and $\gamma\to\infty$ in the first series of numerical experiments and $\sigma=0.05$ and $\gamma=50$ in the other one. We set the number of particles $N=20$ just like in one of the experimental configurations in the strong interaction regime~\cite{Kao_2020}.

As we have mentioned earlier, one can think that above a certain value $\Delta\phi_\mathrm{max}$, the droplet will not coexist with a soliton, but split into two. In such a case, the fidelity should drop rapidly in the vicinity of $\Delta\phi_\mathrm{max}$. Droplet splitting is shown in Ref.~\cite{DePalo_2022}, when a droplet is released from an external confinement and in Ref.~\cite{Edmonds_2020} as a result of an interaction quench. 

Nevertheless, as we can see in Fig.~\ref{fig:fidelities}, no such behaviour is present there. Obviously, the fidelities are smaller in the case with finite both DDI range and contact interaction strength as compared to the other case.

The inset of Fig.~\ref{fig:fidelities} shows us the time evolution of the droplet. 
We imprint a phase difference of $\Delta\phi=\pi/2$ on a quarter of the droplet. From this moment on, we can observe a dark soliton moving with a relative velocity $\beta\approx 0.37$ and accompanied by shock waves.

\section{Conclusions}

All in all, the results shown in this Letter corroborate both the existence of dark solitons in the dipolar Bose gases with strong contact interactions and the possibility of quantum droplet-dark soliton coexistence. 

We put our focus on the strong contact interaction regime, where the quantum droplets emerge in quasi-1D. In this regime, where the system can be modeled with the nonlocal LLGPE,  we deal with a competence between two different types of nonlinearities, the %quadratic
quintic and the cubic one. 
In the limit of infinite contact interaction strength ($\gamma\to\infty$) and zero-range dipolar interactions ($\sigma=0$),
we found an analytical solution for the dark solitons given by Eqs.~\eqref{eq:solutions}. The motionless soliton width diverges when $\gamma_\mathrm{dd}=\frac{2}{3}\pi^2$. As a consequence, the solitons will be ultrawide and easy to observe experimentally in large quasi-1D systems.

We show that in the droplet regime, due to the interplay between different nonlinearities, the soliton exhibits anomalous behaviour -- there exists a gray, but a motionless one. The anomaly is also apparent in the soliton dispersion relation, which contains an additional subbranch.

We complement our analytical considerations with the numerical simulation of an experimental procedure used to generate solitons, i.e. the phase imprinting. We showed the procedure  causes the formation of a dark soliton on top of the droplet, even for large phase jumps, finite $\gamma$ and $\sigma$. This fact disfavours the idea that the phase-imprinting method can lead to an instantaneous droplet splitting.

\section{Acknowledgements}
Center for Theoretical Physics of the Polish Academy of Sciences is a member of the National Laboratory of Atomic, Molecular and Optical Physics (KL FAMO).

J.K., M.{\L}., and K.P. acknowledge support from the (Polish) National Science Center Grant No. 2019/34/E/ST2/00289. W.G. was supported by the Foundation for Polish Science (FNP) via the START scholarship.

J.K. performed the numerical and analytical calculations with help from M.{\L}. (linearization) and W.G. (energy renormalization). K.P. designed and supervised the project. J.K. wrote the manuscript with input of all authors.
\bibliography{apssamp}

\newpage
\renewcommand{\theequation}{S\arabic{equation}}
\renewcommand{\thefigure}{S\arabic{figure}}
\setcounter{equation}{0}
\setcounter{figure}{0}
\section{Supplemental material for:\\Ultrawide dark solitons\\and droplet-soliton coexistence\\in a dipolar Bose gas\\with strong contact interactions}

\subsection{Solitonic solution derivation}

We plug the function $\Phi_s(x,t)\equiv\sqrt{\rho(\zeta)}e^{i\phi(\zeta)}\exp(-i\mu t/\hbar)$ into Eq.~\eqref{eq:Malomed_Salerno_eqn}. The parameter $v$ is the soliton velocity and $\zeta=x-vt$ is the comoving coordinate.
We get a complex equation
\begin{equation}
\begin{split}
\mu\sro\underline{-i\hbar v(\sro)'}+\hbar v\phi'\sro=\\-\frac{\hbar^2}{2m}(\sro)''\underline{-i\frac{\hbar^2}{m}\phi'(\sro)'-i\frac{\hbar^2}{2m}\phi''\sro}\\
+\frac{\hbar^2}{2m}(\phi')^2\sro+\frac{\hbar^2\pi^2}{2m}\sro^5-\gdd\sro^3,
\end{split}
\end{equation}
which can be split into the real and \underline{imaginary} parts.

We begin with the imaginary part, which after a few steps takes the following form:
\begin{equation}
    \left(\frac{\hbar}{m}\phi'\rho-v\rho\right)'=0
\end{equation}
and we further integrate it with an integration constant $Q$:
\begin{equation}
\phi'=\frac{m}{\hbar}\left(v+\frac{Q}{\rho}\right).
\label{eq:phi_prime}
\end{equation}

Then, we can move to the real part and, similarily, we transorm it to
\begin{equation}
\begin{split}
  \left(\mu+\frac{mv^2}{2}\right)\sro(\sro)'+\frac{\hbar^2}{2m}(\sro)''(\sro)'-\frac{mQ^2}{2}\\
  \times\frac{(\sro)'}{\rho\sro}-\frac{\hbar^2\pi^2}{2m}\sro^5(\sro)'+\gdd\sro^3(\sro)'=0.
\end{split}
\end{equation}
Eventually, we get
\begin{equation}
\begin{split}
    2\left(\mu+\frac{mv^2}{2}\right)\rho^2+\frac{\hbar^2}{m}\left(\frac{\rho'}{2}\right)^2\\
    +mQ^2-\frac{\hbar^2\pi^2}{3m}\rho^4+\gdd\rho^3-4V\rho=0.
    \label{eq:dens_prime}
\end{split}
\end{equation}

When we impose the conditions that far from the soliton both the density and phase are constant, i.e. $\lim_{\zeta\to\pm\infty}\rho(\zeta)=\rho_\infty$ and $\lim_{\zeta\to\pm\infty}\phi(\zeta)=\phi_{\pm\infty}$, we straightforwardly get the integration constant values:
\begin{subequations}
\begin{equation}
Q=-v\rho_\infty,
\end{equation}
\begin{equation}
    V=\frac{\hbar^2\pi^2\rho_\infty^3}{6m}-\frac{\gdd}{4}\rho_\infty^2+\frac{1}{2}mv^2\rho_\infty.
\end{equation}
\end{subequations}

Now, we are able to rewrite Eqs.~\eqref{eq:dens_prime} and~\eqref{eq:phi_prime} in the following forms:
\begin{subequations}
\begin{equation}
    \int_0^\zeta d\sigma=\int_{\rho_\mathrm{min}}^{\rho(\zeta)}\frac{d\eta}{2\sqrt{-\frac{m}{\hbar^2}U(\rho)}},
    \label{eq:density_integral}
\end{equation}
with the polynomial $U(\rho)=-\frac{\hbar^2\pi^2}{3m}(\rho-\rho_\infty)^2\left[\rho^2+\left(2\rho_\infty-\frac{3m\gdd}{\hbar^2\pi^2}\right)\rho-\frac{3m^2v^2}{\hbar^2\pi^2}\right]$.
\begin{equation}
\begin{split}
    \phi(\zeta)=\frac{mv}{\hbar}\int_0^\zeta\left(1-\frac{\rho_\infty}{\rho(x)}\right)dx=\\
    -\frac{mv(D+1)(\rho_\infty-\rho_\mathrm{min})}{\hbar DW\rho_\infty}\int_0^{W\zeta} \frac{du}{a+\cosh(u)},
   \label{eq:phase_integral}
\end{split}
\end{equation}
\end{subequations}
where $a\equiv \frac{\rho_\mathrm{min}}{\rho_\infty}+\frac{\rho_\mathrm{min}}{D\rho_\infty}-1$, $D\equiv\frac{\rho_\mathrm{min}-\rho_1}{2\rho_\infty-\rho_1-\rho_\mathrm{min}}$, and $W\equiv 2\sqrt{\frac{\pi^2}{3}(\rho_\infty-\rho_\mathrm{min})(\rho_\infty-\rho_1)}$.
 Having integrated Eqs.~\eqref{eq:density_integral} and~\eqref{eq:phase_integral}, we obtain Eqs.~\eqref{eq:density_sol} and~\eqref{eq:phase_sol} respectively.

\subsection{Numerical codes}
The LLGPE with the nonlocal term is a complex, non-linear partial differential equation.  The function $\psi(x)$ discretized on a spatial mesh with $N_x$ fixed points with lattice constant $DX=\frac{L}{N_x}$. $L$ is the box size, we assume periodic boundary conditions $\psi(-\frac{L}{2})=\psi(\frac{L}{2})$.  The real-time evolution is done with the use of the split-step numerical method. The evolution in both the kinetic and dipolar (if the dipolar range $\sigma>0$; we use the spatial domain otherwise) is done in the momentum domain, whereas the contact interaction term is calculated in the spatial domain. No external potential is used. The initial droplet state is obtain with the imaginary-time evolution when $t\to -i\tau$.

The program written in C++ implementing the algorithm above is available here: \url{https://gitlab.com/jakkop/mudge/-/releases/v30Jun2022}. The program uses the W-DATA format dedicated to store data in numerical experiments with ultracold Bose and Fermi gases. The W-DATA project is a part of the W-SLDA toolkit~\cite{WSLDAToolkit}.

The Python code and data needed to reproduce all the figures are available at request to the corresponding author.

We use the following approximations of $e_\mathrm{LL}(\gamma)$ for $\gamma<1$:
\begin{equation}
\begin{split}
    e(\gamma)\approx\gamma-\frac{4}{3 \pi} \gamma ^{3/2} +\frac{\pi ^2-6}{6 \pi ^2} \gamma^2 - \frac{4-3 \zeta (3)}{8 \pi ^3}\gamma^{5/2}\\ - \frac{4-3 \zeta (3)}{24 \pi^4} \gamma^3 - \frac{45 \zeta(5)-60 \zeta (3)+32}{1024 \pi ^5} \gamma^{7/2}\\ - \frac{3 \left[15 \zeta(5)-4 \zeta (3)-6\zeta (3)^2\right]}{2048 \pi ^6}\gamma^4\\ -\frac{8505\zeta(7)-2520\zeta(5)+4368\zeta(3)-6048\zeta(3)^2 -1024}{786432 \pi^7}\gamma^{9/2}\\
-\frac{9[273\zeta(7)-120\zeta(5)+16\zeta(3)-120\zeta(3)\zeta(5)]} {131072 \pi ^8}\gamma^5
\end{split}
\end{equation}
and 
\begin{equation}
\begin{split}
    e_{\rm LL}(\gamma)\approx \frac{\pi^2}{3}\Bigg(1-\frac{4}{\gamma}+\frac{12}{\gamma^2}-
\frac{10.9448}{\gamma^3}-\frac{130.552}{\gamma^4}+\frac{804.13}{\gamma^5}\\-\frac{910.345}{\gamma^6}-\frac{15423.8}{\gamma^7}+
\frac{100559.}{\gamma^8}-\frac{67110.5}{\gamma^9}\\
-\frac{2.64681 \times 10^6}{\gamma^{10}}+\frac{1.55627 \times 10^7}{\gamma^{11}}+\frac{4.69185 \times 10^6}{\gamma^{12}}\\-
\frac{5.35057 \times 10^8}{\gamma^{13}}+\frac{2.6096 \times 10^9}{\gamma^{14}}+\frac{4.84076 \times 10^9}{\gamma^{15}}\\
-\frac{1.16548 \times 10^{11}}{\gamma^{16}}+
\frac{4.35667 \times 10^{11}}{\gamma^{17}}+\frac{1.93421 \times 10^{12}}{\gamma^{18}}\\-\frac{2.60894 \times 10^{13}}{\gamma^{19}}+\frac{6.51416 \times 10^{13}}{\gamma^{20}} \Bigg).
\end{split}
\end{equation}
for $\gamma\geq1$.

\section{Energy renormalization}

\begin{figure}
    \centering
    \includegraphics[width=0.5\textwidth]{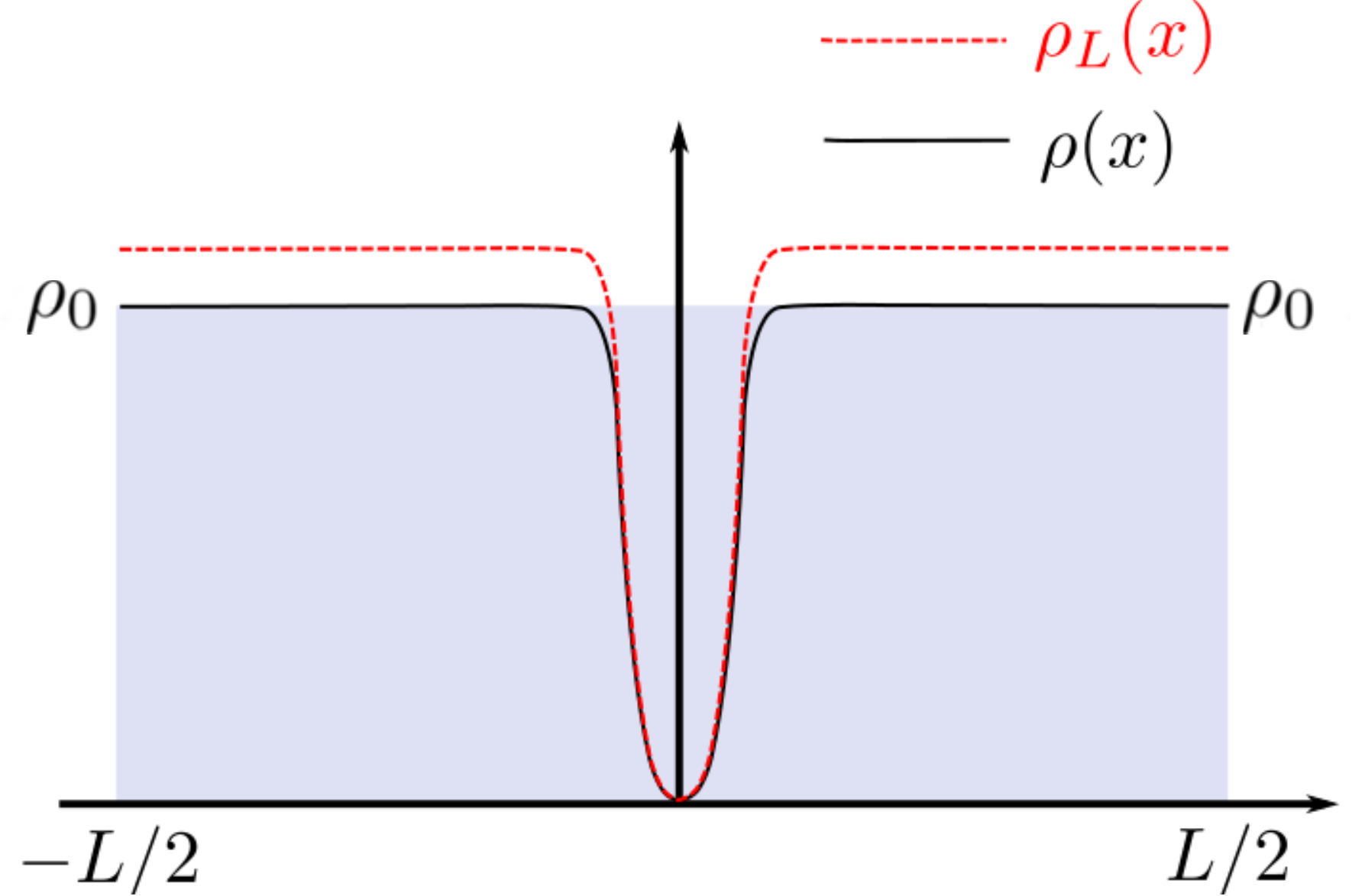}
    \caption{Sketch illustrating the auxillary function needed for the proper renormalization of the   energy of a soliton. }
    \label{fig:suppl1}
\end{figure}
We present here the standard renormalization procedure and the final formulae for the  solitonic excitation propagating on an infinitely long line.
In such a system, the energy of a homogeneous gas is infinitely large, and so is the energy of the gas with a soliton. 
Therefore, the energy of a solitonic excitation, denoted with the letter $\mathcal{E}$ in the main text, is defined as difference between these two energies.
However, firstly one has to compute the difference in a finite box of the length $L$ and then evaluate it in the thermodynamic limit ($N\to\infty$ and $L\to\infty$ with $N/L=\rho_0= const$).
For the finite box, it is necessary to account for the increase of the density outside the soliton due to the particles shifted from the solitonic dip, as depcited with $\rho_{\rm L}$ in  Fig.~\ref{fig:suppl1}. 

The energy function, conserved during motion defined in Eq.~\eqref{eq:Malomed_Salerno_eqn} of the main text is given by 
\begin{equation}
E^L(\rho) =   \int_{-L/2}^{L/2}\, \left( \frac{\hbar^2\pi^2}{6m} \rho^3 - \frac{g_{\rm dd}}{2} \rho^2 \right)\,dx
\end{equation}

The reference energy is the energy of the homogeneous gas with the density $\rho_{0}=N/L$:
\begin{equation}
E^L\left(\rho_0  \right) =\frac{\hbar^2\pi^2}{6m} \frac{N^3}{L^2} - \frac{g_{\rm dd}}{2} \frac{N^2}{L}.
\end{equation}

In the main text, for the solitonic solution $\rho(x)$ we have set $\rho_\infty=\rho_0$, which is true in thermodynamic limit. For finite $L$, however, it leads to a situation, where $\rho(x)$ corresponds to a smaller number of atoms than in the homogeneous reference system. It makes it impossible to compare the energies of these two states in a reasonable way. To solve this problem, we need to normalize $\rho(x)$ by a proper factor:
\begin{multline}
\rho_L(x) = \rho(x)\frac{N}{\int_{-L/2}^{L/2}\rho(x')dx'}\\=\rho(x)\frac{N}{N+\int_{-L/2}^{L/2}\left(\rho(x')-\rho_0\right)dx'}
\end{multline}
such that $\int_{-L/2}^{L/2}\,\rho_L(x)\,dx=N$.
Note that in the thermodynamic limit $\rho_L(x)$ converges to $\rho(x)$, as $\int_{-L/2}^{L/2}\left(\rho(x')-\rho_0\right)dx'$ converges to constant.

The solitonic excitation energy in this box equals
\begin{equation}
\mathcal{E}^L=E^L\left(\rho_L\right)-E^L\left(\rho_0\right)
\end{equation}
In the thermodynamic limit one gets
\begin{equation}
    \mathcal{E} = \lim_{\substack{L\to\infty\\
    N\to\infty\\ N/L=\rho_{0}}}\mathcal{E}^L = \mathcal{E}_{\rm int}+ \mathcal{E}_{\rm dip},
\end{equation}

where
\begin{eqnarray}
\mathcal{E}_{\rm int} &=&\frac{\hbar^2\pi^2}{6 m} \int_{-\infty}^{\infty} \left(3\rho_0 w^2(x) + w^3(x)\right)\,dx \\
\mathcal{E}_{\rm dd} &=& -\frac{\gdd}{2} \int_{-\infty}^{\infty}  w^2(x)\, dx,\\
w(x) &\equiv& \rho(x) - \rho_{0}
\end{eqnarray}

In our case, the renormalized  self-interaction $\mathcal{E}_\mathrm{int}$, dipolar $\mathcal{E}_\mathrm{dd}$ energy terms and the renormalized momentum $\mathcal{P}$ (see Ref.~\cite{Kivshar_1998}) are given by Eqs.~\eqref{eq:energy_renorm}-\eqref{eq:momentum_renorm}.
\begin{widetext}
\begin{subequations}
\label{eq:energy_renorm}
\begin{equation}
\begin{split}
\mathcal{E}_\mathrm{int}=
\frac{\hbar^2\pi^2}{6m}\left[\frac{-3\rho_\infty(\rho_\infty-\rho_\mathrm{min})^2(1+D)^2}{WD^2}\frac{2\sqrt{D^{-2}-1}+D^{-1}\ln[2D^{-2}-2D^{-1}\sqrt{D^{-2}-1}-1]}{\sqrt{D^{-2}-1}^3}\right.+\\
    \left.+\frac{(\rho_\infty-\rho_\mathrm{min})^3(1+D)^3}{WD^3}\frac{6D^{-1}\sqrt{D^{-2}-1}+(2D^{-2}+1)\ln[2D^{-2}-2D^{-1}\sqrt{D^{-2}-1}-1]}{2\sqrt{D^{-2}-1}^5}\right],
 \end{split}
\end{equation}
\begin{equation}
\begin{split}
    \mathcal{E}_\mathrm{dd}=
    \frac{\gdd(\rho_\infty-\rho_\mathrm{min})^2(1+D)^2}{WD^2}
    \frac{2\sqrt{D^{-2}-1}+D^{-1}\ln[2D^{-2}-2D^{-1}\sqrt{D^{-2}-1}-1]}{2\sqrt{D^{-2}-1}^3},
\end{split}
\end{equation}
\end{subequations}
\begin{equation}
\label{eq:momentum_renorm}
\begin{split}
    \mathcal{P}=\frac{i\hbar}{2}\lim_{L\to\infty}\int_{-\frac{L}{2}}^{\frac{L}{2}}\left[\psi(x)\frac{d\psi^*(x)}{dx}-\psi^*(x)\frac{d\psi(x)}{dx}\right]dx-2\hbar\rho_\infty\phi_\infty=mv\frac{(\rho_\infty-\rho_\mathrm{min})(1+D)}{WD\sqrt{D^{-2}-1}}\\
    \times\ln[2D^{-2}-2D^{-1}\sqrt{D^{-2}-1}-1]-2\hbar\rho_\infty\phi_\infty,
\end{split}
\end{equation}
\end{widetext}
with $
    2\phi_\infty=\frac{4mv(D+1)(\frac{\rho_\mathrm{min}}{\rho_\infty}-1)}{\hbar DW\sqrt{1-a^2}}\arctan\left(\frac{a-1}{\sqrt{1-a^2}}\right)$.
\section{Phase imprinting and fidelity calculation}
We prepare the initial droplet state $\psi_\mathrm{num}(x,t=0)$ in the course of imaginary-time evolution. Then, we evolve it in the real time and imprint the phase pattern $\phi_\mathrm{num}(x,t_0)=\phi_\mathrm{num}(x,t_0-dt)+\Delta\phi\,\theta_H(x-W_D/4)$ at certain time $t_0$, where  $\theta_H$ is the Heaviside theta function and $W_D$ is the droplet width.

We choose fidelity $F(\psi_1,\psi_2)=|\langle \psi_1|\psi_2\rangle|^2$ as the measure of the similarity between the state imprinted on the droplet and the analytical solution given by Eqs.~\eqref{eq:solutions}.

Phase imprinting generates a soliton accompanied by shock waves and numerical noise. Thus, we need to establish a numerical procedure in such way that the measurement of fidelity is the least possibly affected by the presence of the two.
\begin{enumerate}
\item We choose the measurement time $\tilde t$ in the very moment when the shock wave escapes the droplet on the right ($\tilde t\approx t_0+\frac{W_D/4}{2c}$). 
\item We use a moving average on the density profile $|\psi_\mathrm{num}(x,\tilde t)|^2$, get a smoothened density $\bar\rho(x)$ and find its minimum within the droplet bulk $\bar\rho_\mathrm{min}=\min_{-\frac{W_D}{2}<x<\frac{W_D}{2}}\bar\rho(x)=\bar\rho(x_\mathrm{min})$.
\item We choose the bounding points $x_L=\min \{x: \forall_{x<x_\mathrm{min}}\bar\rho(x)\leqslant\bar\rho(x_\mathrm{min})+\frac{3}{4}[|\psi_\mathrm{num}(x,t=0)|^2-\bar\rho(x_\mathrm{min})]\}$ and $x_R= \max \{x: \forall_{x>x_\mathrm{min}}\bar\rho(x)\leqslant\bar\rho(x_\mathrm{min})+\frac{3}{4}[|\psi_\mathrm{num}(x,t=0)|^2-\bar\rho(x_\mathrm{min})]\}$.
\item For a given $\frac{\rho_\mathrm{eq}}{\rho_0}\gamma_\mathrm{dd}=\frac{m}{\hbar^2}\frac{\gdd}{|\Phi_\mathrm{MS}(x=0)|}=\frac{2}{3}\pi^2\left[1-\mathrm{sech}\left(\sqrt{\frac{N}{3}}\pi\right)\right]^{-1}$, with $N$ being the particle number, we choose a relative soliton velocity $\beta$ such that $\bar\rho(x_L\leqslant x\leqslant x_R)$ and $|\psi(x_L\leqslant x-x_\mathrm{min}\leqslant x_R)|^2$ given by Eqs.~\eqref{eq:solutions}. One can perform an a posteriori check and see if the fitted velocity agrees with the soliton velocity just like in Fig.~4.

 \item Eventually, we perform the fidelity measurement:
\begin{equation}
F=\frac{\left|\int_{x_L}^{x_R}\psi_\mathrm{num}^*(x,\tilde t)\psi(x-x_\mathrm{min})dx\right|^2}{\int_{x_L}^{x_R}|\psi_\mathrm{num}(x,\tilde t)|^2dx \int_{x_L}^{x_R}|\psi(x-x_\mathrm{min})|^2dx}.
\end{equation}
\end{enumerate}
We provide films with 4 chosen cases of soliton imprinting:
\begin{itemize}
\item Film 1: \texttt{Phase-imprinted soliton moving in a quantum droplet ($\Delta\phi=\pi/2$, $\sigma=0$)}\\
Parameters: effective dipolar range $\sigma=0$, contact interaction strength $\gamma\to\infty$, dipolar interaction strength $\gamma_\mathrm{dd}=9$, phase imprint $\Delta \phi=\pi/2$\\
Link: \url{https://youtu.be/MTVCQ6YbfCM}
\item Film 2: \texttt{Phase-imprinted soliton moving in a quantum droplet ($\Delta\phi=\pi/4$, $\sigma=0$)}\\
Parameters: effective dipolar range $\sigma=0$, contact interaction strength $\gamma\to\infty$, dipolar interaction strength $\gamma_\mathrm{dd}=10$, phase imprint $\Delta \phi=\pi/4$\\
Link: \url{https://youtu.be/WNkGKHw0d4o}
\item Film 3: \texttt{Phase-imprinted soliton moving in a quantum droplet ($\Delta\phi=\pi/2$, $\sigma=0.05$)}\\
Parameters: effective dipolar range $\sigma=0.05$, contact interaction strength $\gamma\to\infty$, dipolar interaction strength $\gamma_\mathrm{dd}=9$, phase imprint $\Delta \phi=\pi/2$\\
Link: \url{https://youtu.be/JcywR4GDiRE} 
\item Film 4: \texttt{Phase-imprinted soliton moving in a quantum droplet ($\Delta\phi=\pi/4$, $\sigma=0.05$)}\\
Parameters: effective dipolar range $\sigma=0.05$, contact interaction strength $\gamma\to\infty$,  dipolar interaction strength $\gamma_\mathrm{dd}=10$, phase imprint $\Delta \phi=\pi/4$\\
Link: \url{https://youtu.be/74YTfdXjULA}
\end{itemize}
\end{document}